\documentclass[]{aa}
\usepackage{graphicx}
\usepackage{txfonts}

\newcommand{\be}{\begin{equation}}
\newcommand{\ee}{\end{equation}}
\newcommand{\bea}{\begin{eqnarray}}
\newcommand{\eea}{\end{eqnarray}}
\newcommand{\eq}[1]{ \begin{equation} #1 \end{equation} }

\begin{document}

\title{Stellar scattering and the origin of the planet around $\gamma$-Cephei-A}

\author{J.G. Mart\'{\i}  \and C. Beaug\'e}

\institute{Instituto de Astronom\'ia Te\'orica y Experimental, Observatorio Astron\'omico, \\ 
Universidad Nacional de C\'ordoba, Laprida 854, (X5000BGR) C\'ordoba, Argentina}

\abstract{In the last years several exoplanets have been discovered that orbit one component of a compact binary system (separation $< 50$ astronomical units), the probably best-known case is $\gamma$-Cephei. So far, all attempts to explain the in-situ formation of these planets has been unsuccessful, in part because of the strong gravitational perturbations of the secondary star on any initial planetesimal swarm.}
{Here we test whether planetary bodies in compact binaries, in particular $\gamma$-Cephei, could have originated from a close encounter with a passing star, assuming initial configurations for the stellar system suitable for planetary formation. In other words, we analyze whether the orbital configuration of the current binary system might have been generated after the formation of the planet, and as a consequence of a close encounter with a third star in hyperbolic orbit.}
{We performed a series of time-reverse N-body simulations of stellar scattering events in which the present-day configuration of $\gamma$-Cephei was used as the initial condition plus a hypothetical third star as an impactor. We analyzed which configurations and system parameters could have given birth to the current system.}
{Depending on the maximum impact velocity allowed for accretional collisions, we find that between $\sim 1 \%$ and $\sim 5 \%$ of stellar encounters correspond to an ``original'' system in which planetary formation around the primary star is not inhibited by the secondary, but is acceptable within the classical core-accretion scenario. Thus, although not highly probable, it is plausible that stellar encounters may have played a significant role in shaping these types of exoplanetary systems.}{}

\keywords{planets and satellites: formation - stars: binaries: $\gamma$-Cephei.}

\date{}

\maketitle

\section{Introduction}

Studies of planetary formation in binary stellar systems are an intriguing and challenging task. Although the majority of stars are members of such systems (Duquennoy \& Mayor 1991), only a handful of exoplanets have so far been detected around compact binaries (i.e. stellar separation below $\sim 50$ AU). Even so, their very existence appears to be inconsistent with models and theories that work for single stars, which pushes some crucial parameters to extreme values.

The case of the $\gamma$-Cephei binary is particularly interesting because it is one of the most extreme systems that harbors a giant planet. The stellar system is formed by a primary star with a mass of $M_{A} = 1.6 M_{\odot}$ and a secondary with a mass of $M_{B} = 0.34 M_{\odot}$. The $M_{B}$ body orbits around $M_{A}$ with a semimajor axis of $a_{AB} = 18.5 \; \textrm{AU}$ and an eccentricity of $e_{AB} = 0.36$ (Hatzes et al. 2003).
 
The giant planet that orbits $M_{A}$ has been the object of many studies over the past decade (e.g. Marzari \& Scholl 2002; Th\'ebault 2004; Th\'ebault 2006; Haghighipour \& Raymond 2007; Paardekooper et al. 2008; Beaug\'e et al. 2010; etc.) and we have yet to find a suitable physical scenario to explain its existence. 
Artymowicz and Lubow (1994) showed that the circumplanetary disk is tidally truncated by its companion. This truncation occurs at a location close to the outer limit for dynamical stability of solid particles, and therefore safely lies beyond the position of all detected exoplanets. It nevertheless poses a problem, especially for giant planet formation, because it deprives the disk from a large fraction of its mass. Another consequence of disk truncation is that it shortens the lifetime of the disk, and in turn the timespan for gaseous planet formation (Cieza et al. 2009). To complicate the scenario even more, Nelson (2000) suggested that for an equal-mass binary of separation $50$ AU, temperatures in the disk might stay too high to allow grains to condense.

Even if grains can condense, binary perturbations might impede their growth by increasing the impact velocities beyond the disruption limit (e.g. Marzari \& Scholl 2000). Although gas drag helps to introduce a phase alignment in the pericenter and, thus, to decrease the relative velocities of nearby bodies, this appears to work only for equal-size planetesimals (e.g. Thebault et al. 2006) and circular gas disks. Eccentric and even precessing gas disks complicate the picture still more (Paardekooper et al. 2009; Marzari et al. 2009; Marzari et al. 2012), although it may indeed work in our favor in the outer parts of the disk (Beaug\'e et al. 2010). 

In recent years several studies have been performed on the role of mutual inclinations between the planetesimals, the gas disk, and the binary companion (Marzari et al. 2009b; Xie et al. 2010; Thebault et al. 2010; M\"uller \& Kley 2012; Zhao et al. 2012.). However, even this additional degree of freedom does not appear to be a solution and, in the particular case of $\gamma$-Cephei, does not seem to favor planetary formation for bodies at $a > 2$ AU, which is expected to be the breading ground for the planet detected in this system. 

In a recent review of the state of the art in this problem, Thebault (2011) wondered whether the whole idea of core-instability planetary formation is feasible in such systems, or if we must resort to gas-instability models
(Duchene 2010). However, as Thebault (2011) points out, ``the disk instability scenario also encounters severe difficulties in the context of close binaries, and it is too early to know if it can be a viable alternative formation channel''.

So, the question remains: How did the planet manage to form around the star $\gamma$-Cephei-A?
Here we {\bf revisit} an alternative route, based on the concept of stellar scattering. We assume that planet $m_{p}$ formed (by one of the accepted models of planetary formation) around $M_{A}$, but in a different dynamical configuration in which the star $M_{A}$ is single or the companion $M_{B}$ is sufficiently wide to allow the formation of the planetary system.

The scenario explained is not new. Pfahl (2005) and Portegies Zwart \& McMillan (2005) discussed such a mechanism to explain the origin of a planet then believed to exist in the HD188753 system (Konacki 2005), although the existence of this body was subsequently questioned and is so far unconfirmed. 

A slightly different analysis was later presented by Marzari \& Barbieri (2007a, 2007b), were the authors studied the survival of circumstellar planets in dynamically unstable triple stellar systems or in binaries that suffered a close encounter with a third (background) star. These papers explored, for the first time the possibility that the current configuration of the $\gamma$-Cephei stellar components may not be primordial, and that the exoplanet may have been formed under different circumstances. 

In this paper we again address the proposal of Marzari \& Barbieri (2007a), and study the effects of hyperbolic fly-bys of background stars in a binary system similar to $\gamma$-Cephei. The idea is to see under which initial conditions and system parameters it is possible to transform a primordial accretion-friendly stellar system to the current (accretion-hostile) configuration. Since our is dedicated to a single binary system, we are able to explore a larger set of parameters and situations, as well as different types of outcomes. We also include a statistical analysis of the results considering distribution probabilities for impacting velocities and stellar masses.

The work is organized as follows. In Section 2 we review the idea of gravitational scattering, with special emphasis on hierarchical three-body systems. Section 3 presents our N-body code and some test runs to analyze the importance of the free parameters in the scattering outcomes. Our main numerical simulations are discussed in Section 4, while a probability analysis of positive outcomes is presented in Section 5. Conclusions finally close the paper in Section 6.

\section{Gravitational scattering in hierarchical three-body systems}\label{sec2}

The gravitational interaction between a star and a binary system is a classical problem in stellar dynamics (e.g. Heggie 1975; Hills 1975; Hut \& Bahcall 1983; Heggie \& Hut 1993; Portegies Zwart et al. 1997; etc.), although it is focused on the relationship between binaries and the dynamical evolution of open clusters. The same principles, however, may also be applied to other problems, such as the origin of exoplanets in multiple stellar systems (Pfahl 2005, Portegies Zwart \& McMillan 2005) and the dynamical evolution of ``cold'' exoplanets with orbits very distant from the central star (e.g. Veras et al. 2009). 

\subsection{The hyperbolic two-body problem}

We assume a compact stellar binary system with masses $m_A$ and $m_B$, whose orbital separation is small compared with the initial distance of an incoming star $m_C$. In this scenario, and at least for the initial conditions of the hyperbolic encounter, we may approximate the two binary components by a unique body of mass $m_A+m_B$ (see Hut \& Bahcall 1983), and refer the orbit of the impactor with respect to the barycenter of the binary. 

In classical scattering problems, the outcome of the fly-by may be defined by two fundamental parameters: the infall velocity of the incoming body at infinity $v_{\infty} = |\dot{\textbf{r}}(t \to -\infty)|$ and the impact parameter $b$, defined as the minimum distance between the two bodies if there were no deflection. Alternatively, it is possible to substitute $b$ with the minimum distance $r_{min}=a(1-e)$ between the impactor and the center of mass of the pair $m_A+m_B$. 

Following the approach developed by Hills (1975), we can approximate the total orbital energy of the three-star system, prior to the encounter,  by
\eq{ 
E_{\rm tot} = E_{AB} + E_C = -\frac{\kappa}{2a_{AB}} + \frac{1}{2} \mu v_\infty^2,
\label{eq13}
}
where both $\kappa$ and $\mu$ are two mass factors defined according to
\eq{ 
\kappa = {\cal G} (m_A + m_B) m_C
\hspace*{0.4cm} ; \hspace*{0.4cm}
\mu = \frac{(m_A + m_B)\, m_C}{m_A + m_B + m_C}
\label{eq12}.
} 
In equation (\ref{eq13}) $E_{AB}$ is the binding energy of the binary and $E_C$ is the kinetic energy of the impactor. $a_{AB}$ is the semimajor axis of $m_B$ with respect to $m_A$. We assume that initially $m_C$ is sufficiently far from the other two that we can neglect its potential energy. The same calculation can also be made after the scattering event has taken place, now in terms of the new semimajor axis of the binary ($a'_{AB}$) and the outgoing velocity of $m_C$ (which we will denote by $v'_\infty$).

We can then use the conservation of $E_{\rm tot}$ to write (see Heggie 1975)
\eq{ 
\frac{1}{2} \mu {v'}_\infty^2  = \frac{1}{2} \mu v_\infty^2 + \Delta E_{AB},
\label{eq14}
}
where $\Delta E_{AB}$ is the change in the binding energy of the binary. The minimum velocity
$v_c$ of the impactor necessary to disrupt the binary is such that $\Delta E_{AB} = E_{AB}$ or, equivalently, 
$E_{\rm tot} = 0$. This gives (Hut \& Bahcall 1983)
\eq{ 
v_c^2  = \frac{\kappa}{\mu} \frac{1}{a_{AB}} 
       = \frac{{\cal G} m_A m_B (m_A + m_B + m_C)}{(m_A+m_B)m_C} \frac{1}{a_{AB}} .
\label{eq15}
}
It will prove useful to write the incoming velocity of the impactor $v_{\infty}$ in units of $v_c$
\eq{
v^* = \frac{v_{\infty}}{v_c},
\label{eq16}
}
which gives an adimensional expression for the incoming velocity. Furthermore, $v_{\infty}$ can be written in terms of the semimajor axis of the hyperbolic orbit as:
\eq{
v_{\infty}^2 = \frac{{\cal G}(m_A+m_B+m_C)}{a}, 
\label{eq17}
}
where $a$ is the the semimajor axis of $m_C$ with respect to the center of mass of the binary $m_A+m_B$. 
Recalling expression (\ref{eq15}), we may now write the relationship between $a$ and $v^*$ as
\eq{
a = -a_{AB} \frac{m_C(m_A+m_B)}{m_A m_B} \frac{1}{{v^*}^2}.
\label{eq18}
}
Similarly, we can obtain the eccentricity of the hyperbolic orbit of $m_C$ as
\eq{
e = 1 - \frac{r_{min}}{a} .
\label{eq19}
}

Finally, if we define $\epsilon_{0}$ as the value of the eccentric anomaly at the beginning of the simulation (when $r = r_0$), we can calculate the initial position and velocity of $m_C$ using the usual expressions for hyperbolic motion (e.g. Butler 2005). 

\subsection{Complete set of variables for the scattering problem}

The set $(v_*,r_{min},\epsilon_0)$ defines the initial conditions for the impactor $m_C$, with respect to the center of mass of $m_A + m_B$, assuming that the differential gravitational interactions from both components do not significantly perturb the two-body hyperbolic orbit (see Hut \& Bahcall 1983 for a similar approach). However, even if this condition is not satisfied, and the orbit of the incoming star is seriously affected by the two central bodies, the values of $(v_*,r_{min})$ are still indicative of the ``unperturbed'' motion of $m_C$, and therefore may still be referred to as variables of the problem. 

To complete the description of the system, we must give the initial conditions for each of the binary components at a given time. These can be uniquely specified by the set of $m_A$-centric orbital elements of $m_B$: $(a_B,e_B,I_B,\lambda_B,\omega_B,\Omega_B)$, where the inclination $I_B$ is given with respect to the orbital plane of the incoming impactor. Since the value of $\epsilon$ is already a variable of the system, it appears less complicated to specify the value of the mean longitude $\lambda_B$ at the time of maximum approach of the impacter (i.e. $r=r_{min}$). From the two-body solution we can then calculate the value of the mean longitude at the initial time $\epsilon=\epsilon_0$ and calculate the position and velocity vectors of $m_A$ and $m_B$ at the beginning of the simulation. 

As with the initial conditions for $m_C$, it is not expected that the orbits of the binary stars remain constant in time. Accordingly, the values for the orbital elements adopted as initial conditions are also ``unperturbed'' values, in the same way as the impact parameter $b$ is the unperturbed minimum distance between the projectile and the target. 

Last of all, and since our ultimate aim will be to study the exoplanet detected around $\gamma$-Cephei-A, we will also assume that a small planetary mass $m_p$ orbits the central star $m_A$. The initial $m_A$-centric orbit of the planet will be circular and specified by values of $(a_p,I_p,\lambda_p,\Omega_p)$, where $\lambda_p$ is once again given at the time when $r_C=r_{min}$. 

To summarize, we need to set initial values for 16 independent variables in order to get a full description of the scattering event. These are
\begin{eqnarray}
&& (m_A,m_B,m_C,m_p) \;+\; (r_{min},v^*) \;+\; \\
&& (a_B,e_B,I_B,\lambda_B,\omega_B,\Omega_B) \;+\; (a_p,I_p,\lambda_p,\Omega_p) 
\nonumber.
\label{eq21}
\end{eqnarray}
We recall that the $m_A$-centric orbital elements of $m_B$ are assumed to correspond to the point at which the impactor has its maximum approach if the impactor had zero mass. 

Of course, previous knowledge of the dynamical system implies that some variables are already known, such as most of the masses and the main orbital elements. However, this still leaves us with a significant number of degrees of freedom to explore.

\begin{figure*}[th!]  
\centering
\mbox{\includegraphics*[width=16.0cm]{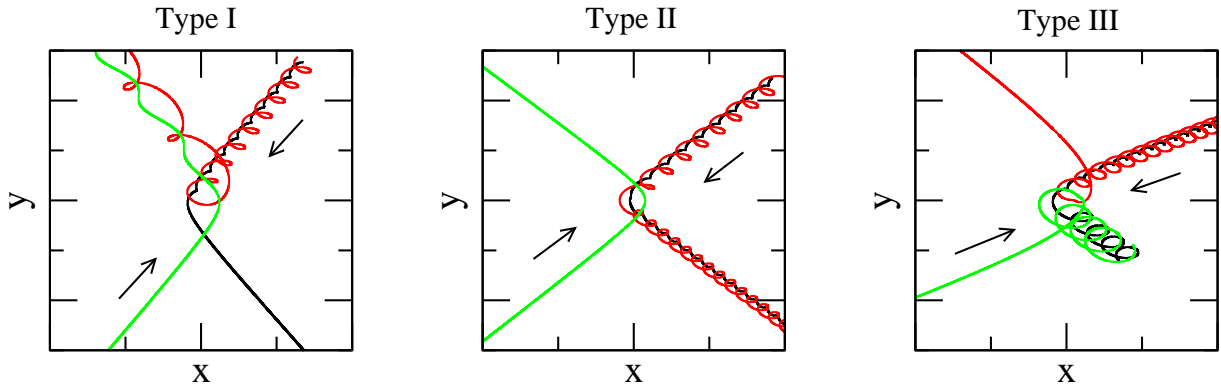}}
\caption{Examples of scattering events that lead to all three types of outcomes. Black curves correspond to $m_A$, red to $m_B$, and light green to the impactor $m_C$.}
\label{fig1}
\end{figure*}

\section{The numerical code}\label{sec3}

Once we set the values for the stellar masses and initial conditions, we integrated the equations of motion using an N-body code to simulate the scattering event. Dynamically, the problem may be characterized as a hierarchical (or nested) full four-body problem, in which all mutual gravitational interactions are taken into account. 

For the simulations described in this work we employed a code constructed around a Bulrisch-Stoer integrator with an error tolerance of $ll=13$. Since the scattering event is expected to be extremely sensitive to the initial conditions, especially if the impact parameter is small, we performed a series of preliminary runs that included back integrations from the final configurations. In all cases there was no significant difference between the two branches of the same trajectory.

Our working hypothesis is that the current $\gamma$-Cephei system could be the outcome of a stellar scattering event that either modified the orbit of the secondary star or caused an exchange of massive bodies. To study whether this is possible, we employes the concept of {\it reverse integration}. In other words, our initial conditions were the present day orbital configuration of the $\gamma$-Cephei system plus an additional star $m_C$ in an arbitrary outward-bound hyperbolic orbit. We then integrated the system backwards in time to reproduce the hypothetical scattering event and find the ``initial conditions'' that led to the present system. It is important to stress that this procedure can only be employed in purely conservative systems where the equations of motion are time-reversible. Therefore, we did not consider the effects of any non-conservative force such as gas drag of tidal effects between the bodies.

We considered that a simulation yields a positive result when the following two conditions are met:

\begin{itemize}

\item The planet $m_p$ remains bounded to $m_A$ in an orbit with moderate to low eccentricity (e.g. $e_p < 0.1$). \\

\item The region around $m_p$ becomes compatible with planetary formation according to standard core-accretion scenario. This could either be a consequence of a significant increase in the semimajor axis of the secondary binary component, or a stellar exchange in which $m_A$ becomes a single star. 

\end{itemize}

Since the equations of motion of the N-body problem are time-reversible, the final outcomes of our simulations would then constitute possible initial conditions that could explain the current planetary system. 

\subsection{Possible outcomes}

Depending on the relative magnitudes between $v_\infty$ and $\Delta E_{AB}$, we can list five possible outcomes of a hyperbolic fly-by: De-excitation, excitation, resonance, ionization, and exchange. The reader is referred to Heggie (1975) for more details, including explicit conditions for each type. 

Not all possible outcomes are relevant for the case at hand. For example, {\it resonance} is a transitory state, at least in the point-mass approximation, and following complex interactions between all three stars, at least one will leave the system in a hyperbolic orbit. {\it Ionization} implies that all three stars acquire unbounded motion with respect to the other two; the original binary is then destroyed, but not replaced. Since in our case we worked with a reverse temporal axis, this would mean that the present-day $\gamma$-Cephei is the outcome of a triple encounter of isolated stars, a highly improbable situation, to say the least. 

Since our aims are more restrictive and focused on finding possible formation mechanisms for $\gamma$-Cephei-A-b, we prefer to redefine the possible outcomes according to what Pfahl (2005) calls ``scenarios''. These are:

\begin{itemize}

\item {\it Type I}: The impactor $m_C$ captures the secondary component $m_B$ during the encounter, leaving behind $m_A$ as a single star with the planet still bounded in a quasi-circular orbit. \\

\item {\it Type II}: While there is no exchange in the memberships, the binary $m_A+m_B$ acquires a sufficiently wide orbit to allow planetary formation in the region around $m_p$. \\

\item {\it Type III}: The  incoming star $m_C$ changes place with $m_B$ where this star is expelled from the system. The new binary $m_A+m_C$ is sufficiently wide to allow planetary formation around the location of $m_p$. 

\end{itemize}

Both types I and III are cataloged as {\it exchange} in Heggie's nomenclature, while type II corresponds to an {\it excitation}. Marzari \& Barbieri (2007a) only discussed type-II outcomes.

In addition to the types listed before, we also found a few cases of three other types of outcomes. Although these were very rare and have only a negligible effect on the statistical analysis performed later on, we will nevertheless include them here for completeness and for a full understanding of the complexity of the problem. These will be referred to as types IV to VI, and correspond to the following scenarios:

\begin{itemize}

\item {\it Type IV}: The impactor $m_C$ not only captures the secondary component $m_B$ during the encounter (which would be cataloged as type I), but also the planet that originally orbited $m_A$. The final configuration then includes a new binary $m_B+m_C$ with the planet orbiting $m_C$.\\

\item {\it Type V}: The impactor $m_C$ carries away both $m_A$ and its planet $m_p$, but in the final configuration the planet now orbits $m_C$. \\

\item {\it Type VI}: The passage of the impactor $m_C$ only captures the planet from the original binary, leaving behind a planet-less $m_A+m_B$ system.

\end{itemize}

As mentioned in the previous section (see equation (\ref{eq12})), we need to specify a total of 16 parameters/variables. Of these, however, several will be fixed to represent the $\gamma$-Cephei system. For the masses we adopted the values
\eq{ 
m_A = 1.6 m_\odot
\hspace*{0.2cm} ; \hspace*{0.2cm}
m_B = 0.4 m_\odot
\hspace*{0.2cm} ; \hspace*{0.2cm}
m_p = 1.7 m_{\rm Jup},
\label{eq22}
}
while for the $m_A$-centric orbit of $m_B$ we assumed:
\eq{ 
a_B = 18.5 \; {\rm AU} 
\hspace*{0.2cm} ; \hspace*{0.2cm}
e_B = 0.36.
\label{eq23}
}
Finally, we assumed for the $m_A$-centric semimajor axis of the planet $m_p$ $a_p=2.1$ AU. The remaining ten parameters where considered to be unknown and free to modify in our series of runs. 

\subsection{Test runs}

Not all variables are created equal, and some are expected to be more influential in determining the outcome of the scattering event. The aim of these preliminary series of runs is to estimate how each parameter affects the dynamics and what may be preferential values. 

We began with three series of runs, each with fixed values of the impactor mass. The values adopted for each run are $m_C=0.2$ (series 1), $m_C=0.4$ (series 2) and $m_C=1.0$ (series 3), with values given in solar masses. All orbits are assumed to lie on the same plane (i.e. $I_i = 0$, and $\Omega_i$ indeterminate) and the remaining angles are taken equal to zero. At the beginning of each integration, the impactor is set at a distance from the binary equal to $r_0=50 Q_B$, where $Q_B = a_B(1+e_B)$ is the apocentric distance between both components. Since the orbital elements of the planet and binary are given at $r=r_{min}$, the outcome of the scattering event is not dependent on the value of $r_0$, as long as it is chosen to be sufficiently large.

Thus, in each series we have only two free parameters left: $(r_{min},v^*)$. For each series we chose a total of $2.5 \times 10^{5}$ values chosen from an equispaced $500 \times 500$ grid within the intervals:
\begin{eqnarray}
1.1 \leq & v_0/v_{esc} & \leq 10  \\ 
0.1 \leq & r_{min}/Q_B & \leq 4  \nonumber ,
\label{eq24}
\end{eqnarray}
where $v_0$ is the magnitude of the initial velocity at $r=r_0$ and $v_{esc}$ the escape velocity at that point. Thus, we consider initial speeds ranging from almost parabolic orbits to highly hyperbolic fly-bys. After some simple algebraic calculations, we find that the corresponding interval in $v^*$ is approximately 
$0.08 \leq v^*\leq 1.7$. 

\begin{figure}[t!]  
\centering
\mbox{\includegraphics*[width=9.0cm]{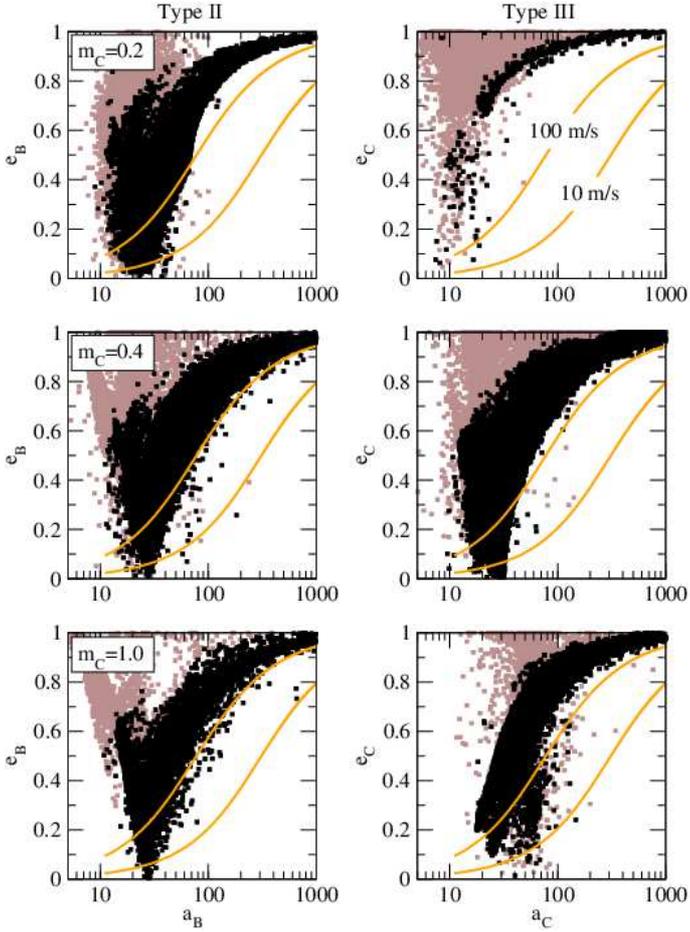}}
\caption{Distribution of the final eccentricity and semimajor axis of the secondary binary with respect to the central star $m_A$, after the scattering event. In the left-hand plots (type II) the secondary star is still $m_B$, while in the right-hand plots it is $m_C$. Brown squares show all runs, while those that preserved the planetary orbit with $e_p < 0.1$ are superimposed in black. The orange lines show level curves of typical impact velocities (see text for details).}
\label{fig2}
\end{figure}

Figure \ref{fig2} shows the post-scattering $m_A$-centric orbital elements of the secondary star of the resulting binary system. Since the central star becomes isolated in type-I interactions, only type II (left) and type III (right) are analyzed. In the first case, $m_B$ remains bounded to $m_A$, so the eccentricity and semimajor axis plotted are those of $m_B$. In type-III outcomes, however, the final binary is now $m_A+m_C$, so the orbital elements are those of $m_C$ with respect to $m_A$. Black squares show those runs in which the planetary orbit remained almost circular, while those in which $m_p$ was severely perturbed are shown in brown. 

In most cases the resulting distribution is V-shaped, roughly delimited on the left by apocentric distances $Q \ge 25$ AU, which corresponds to the apocentric distance of the original binary system. The black squares, on the other hand, are mostly located with $q \ge 12$ AU, which is also the pericentric distance of the original system. It therefore appears that if the new binary has a lower value of $q$, roughly independent of the eccentricity, its gravitational perturbations on the planet would be sufficient to increase its eccentricity beyond the imposed limit $e_p=0.1$. Of course this analysis is qualitative and does not rigorously represent every solution, but it does give an overall picture of the behavior. 

Not all black squares are categorized as positive solutions. As mentioned previously, the final system  must also permit for planetary formation, i.e., allow for an in-situ accretion of $m_p$. Since the present-day $\gamma$-Cephei system is too compact to allow for accretional collisions in the planetesimal swarm (e.g. Th\'ebault et al. 2004; Paardekooper et al. 2008), we must ask about the minimum distance between binary components at which this may be avoided. 

Unfortunately, there is no easy answer to this question, but we may obtain a rough idea from previous numerical studies of planetesimal dynamics in binary systems. Th\'ebault et al. (2006) presented a detailed calculation of the impact velocity of different-sized planetesimals located at $a = 1$ AU in generic binary systems with $m_A=1$ and $m_B = 0.5$, both in units of solar mass. For every point in a grid of values of $(a_B,e_B)$ these authors calculated the average impact velocity $\Delta v$ between a $2.5$ km and a $5$ km solid body, and drew level curves of constant $\Delta v$. According to the authors, accretional collisions are guaranteed for $\Delta v < 10 m/s$, denoting regions where the encounter speeds are practically unaffected by the secondary star. Collisions with $10 < \Delta v < 100 m/s$ are less certain, but may still allow for planetary formation even if some erosion is expected.

Since the impact velocity at any given location is primarily dictated by the forced eccentricity $e_f$ excited by the secondary star on the planetesimal swarm, we can use the results by Th\'ebault et al. (2006) to extrapolate their findings to other binary systems. As determined by Heppenheimer (1978), a first-order expression (in the masses) for the forced eccentricity at a given semimajor axis $a$ may be written as
\eq{ 
e_f = \frac{5}{4} \frac{a}{a_B} \frac{e_B}{1-e_B^2}.
\label{eq25}
}
A second-order expression was calculated recently by Giuppone et al. (2011), which depends explicitly on the stellar masses. However, the difference between both values is only relevant for high values of $e_f$, which is beside the purpose of the present study. 

Using expression (\ref{eq25}) of Th\'ebault's system we can calculate (for our case) the critical value of the forced eccentricity associated to $\Delta v = 10 m/s$ and to $\Delta v = 100 m/s$. We can then use these as reference values and, once again applying (\ref{eq25}), estimate the values of $a_B$ and $e_B$ that give the same value of $e_f$ for a given semimajor axis $a$. Since the forced eccentricity increases with $a$, $a_B$ and $e_B$, in principle we should expect accretional collisions for lower values of all these parameters.

To complete this setup, we must specify a value for $a$. Since the exoplanet detected in $\gamma$-Cephei has a semimajor axis of $2.1$ AU, even an in-situ formation with little planetary migration requires a sufficiently large feeding zone that would extend beyond the snow line to accrete the necessary mass. For the present study, we chose $a=4$ AU; in other words, we required that the entire region up to $4$ AU satisfies the conditions established in Th\'ebault et al. (2006). Although this upper limit may appear rather arbitrary, we consider it as illustrative of the dynamics and not as an actual proposal for the formation process itself.

The orange curves in each frame of Figure \ref{fig2} show the values of the eccentricity and semimajor axis of
the binary system that yield (within the approximation described above) the two values of the impact velocities mentioned before. All binaries located to the right and below each curve satisfy this ad-hoc condition of accretional collision. 

\begin{figure}[t!]  
\centering
\mbox{\includegraphics*[width=9.0cm]{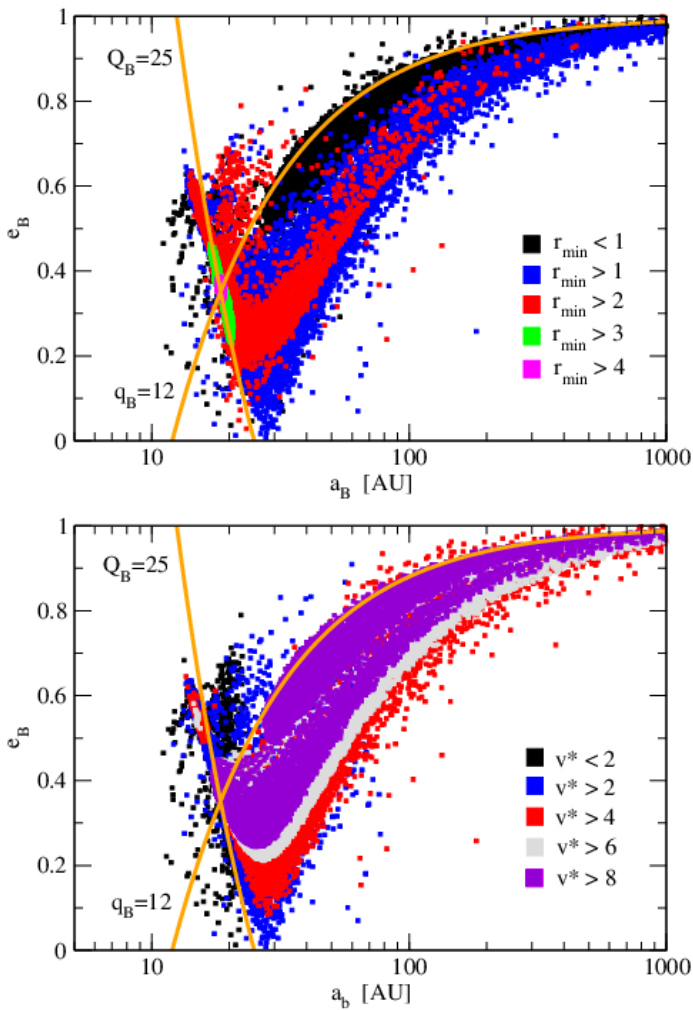}}
\caption{Distribution of the eccentricity and semimajor axis of $m_B$ with respect to $m_A$, separated according to the values of $r_{min}/Q_B$ (top) and $v*$ (bottom). The orange curve on the right corresponds to $q_B=30$ AU.}
\label{fig3}
\end{figure}

Figure \ref{fig3} focuses on the type-II outcomes for $m_C=0.4 M_{\odot}$. The top frame shows the distribution of orbital parameters of the binary as a function of the minimum distance $r_{min}$ of the impactor, in units of $Q_B$. For illustration purposes, we have also drawn the curves of $q_B=a_B(1-e_b)=12$ AU and $Q_B=a_B(1+e_b)=25$ AU, both corresponding to the original values of the system. Thus, the original orbit is located at the intersection point between both curves.

Since distant encounters (i.e. large values of $r_{min}$) imply lower perturbations, the final configuration remains close to the original orbit. However, as the distance decreases, the resulting orbits spread significantly, most of them increasing both $q_B$ and $Q_B$. 

A comparison with the curves of constant $\Delta v$ also shows that the positive solutions occur primarily for impactors whose minimum distance lies between one and three times the apocentric distance of the original binary. Closer encounters are too destructive, while more distant fly-bys are not able to modify the system sufficiently to overcome the planetary formation constraints. 

The lower plot shows the distribution of final orbits as a function of the velocity at infinity of $m_C$, in units of the escape velocity of the system. Low incoming speeds cause a more or less random spread of solutions around the original configuration, mainly as a consequence of the resonance interactions. Higher speeds generate more ordered distributions. Once again, positive solutions are mainly found for intermediate values, between two and six times the escape velocity. 

\begin{table}
\centering
\begin{tabular}{ c | r r r r r }
\hline  \\ [-2ex]
$m_C$ & $0.2$ & $0.4$ & $0.6$ & $0.8$ & $1.0$ \\ [0.5ex]
\hline \\ [-1ex]
  Type I                                             &       --     &   0.5\% &    1.9 \% &    3.3\% &   4.8\%  \\ [1ex]
  Type II   \\ ($\Delta v < 10$ m/s)    &       --     & 0.02\% &  0.01 \% &  0.01\% &  0.01\% \\ [1ex]
  Type II   \\ ($\Delta v < 100$ m/s)  &    1.9\% &   0.9\% &    0.3 \% &    0.2\% &    0.2\% \\ [1ex]
  Type III  \\ ($\Delta v < 10$ m/s)    &      --      &   0.1\% &    0.2 \% &  0.03\% &  0.03\% \\ [1ex]
  Type III  \\ ($\Delta v < 100$ m/s)  &  0.01\% &   1.0\% &    2.1 \% &    0.9\% &    0.2\% \\ [1ex]
\hline
\end{tabular}
\caption{Percentage of positive outcomes per event type and impactor mass.}
\label{table1}
\end{table}

Finally, the percentage of positive results are summarized in Table \ref{table1}, were we have extended the series to include a total of five values of $m_C$ between $0.2 M_{\odot}$ and $1 M_{\odot}$. For most impactor masses, type I positive solutions are the most common because they contain no restrictions to the orbit of the outgoing binary. We reecall that in this type of outcome the central star $m_A$ becomes isolated and the new binary is composed of $m_B+m_C$. All other types of solutions are rare, as already noted by Marzari \& Barbieri (2007a), usually less than $1 \%$ of all scattering events. If we consider the very restrictive $\Delta v < 10$ m/s limit, only a few $0.01 \%$ of the encounters yield final configurations suitable for planetary formation around $m_A$.

\section{Full numerical simulations} \label{sec4}

We now repeated the simulations, this time with random values of all angular variables $(I_B, \lambda_B,\varpi_B,\Omega_B,\lambda_p,\varpi_p)$. In other words, we extended the experiments to the 3D case and considered arbitrary orientations of the orbits of both the binary and the planet. However, in all cases we considered the planet initially in the same orbital plane as the secondary, i.e. $I_p=I_B$ and $\Omega_p=\Omega_B$. One of the aspects we wished to study is the change in the mutual inclination between the planet and the binary perturber after the passage of the impactor. It was already noted in Marzari \& Barbieri (2007ab) that high inclinations between the impactor and the binary may lead to greater disruptions of the system, so these configuration may cause large variations in $I_p$, leaving behind planetary systems with high inclinations with respect to the plane of the binary.

\begin{figure}[t!]  
\centering
\mbox{\includegraphics*[width=9.0cm]{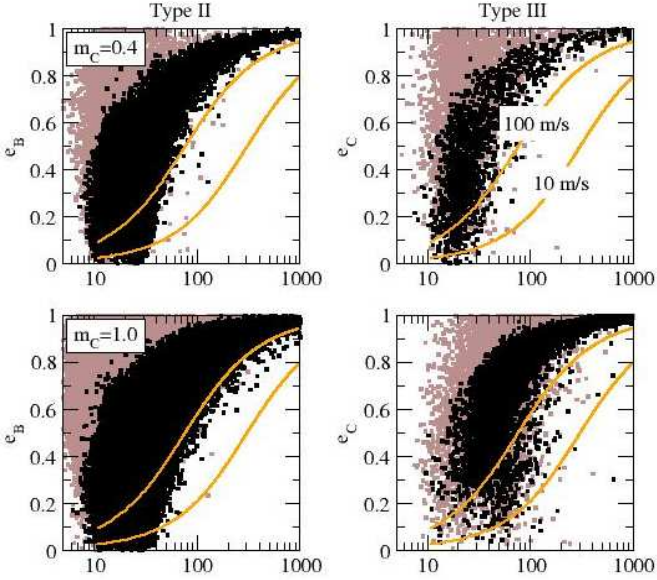}}
\caption{Same as Figure \ref{fig2}, but for random orientations and inclinations of the orbits. Top plots show results for an impactor mass of $m_C=0.4 M_\odot$, while the bottom plots correspond to  $m_C=1 M_\odot$.}
\label{fig4}
\end{figure}

Figure \ref{fig4} shows the eccentricity and semimajor axis distribution of the resulting binaries, where the left-hand plots show the type-II outcomes and the right-hand graphs correspond to type III. As with Figure \ref{fig2}, the symbols in brown show all the outcomes while those in black maintained the planet in an almost circular orbit ($e_p < 0.1$). Compared with Figure \ref{fig2}, we see a greater proportion of positive results for type-II outcomes, but a significant decrease in positive results for type III (see Table \ref{table2}). In fact, the number of Type III outcomes with relative encounter speed $\Delta v < 10$ m/s expected for small planetesimals is less than $0.01 \%$ for all impactor masses. Results also show a serious reduction in the positive outcomes of type-I encounters, which are now typically only half those found in our test runs. 

\begin{figure}[t!]  
\centering
\mbox{\includegraphics*[width=9.0cm]{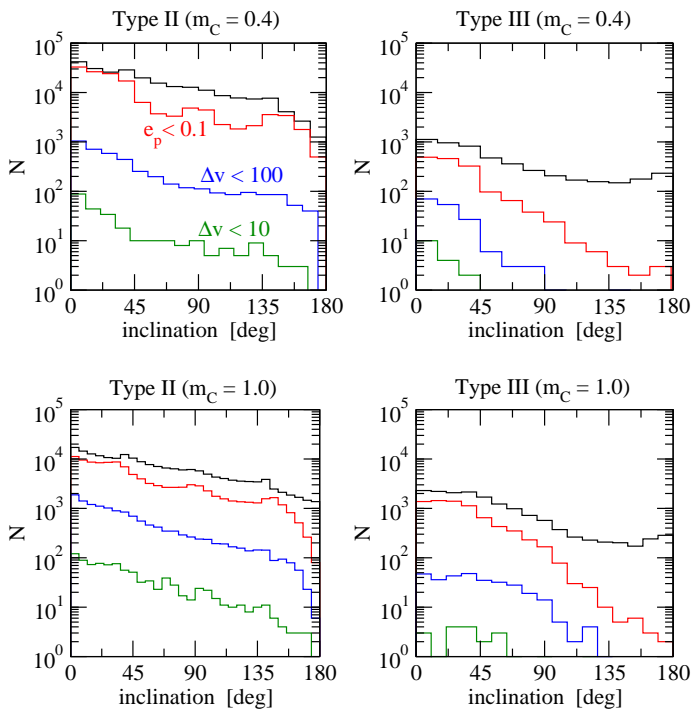}}
\caption{Distribution of the final inclinations of the planetary orbit with respect to the orbit of the binary for the full 3D simulations. Black histograms show all type II (left) and type III (right) outcomes. Those leading to planetary eccentricities below $0.1$ are shown in red, while the accretional friendly final systems are depicted in blue and green.}
\label{fig5}
\end{figure}

Finally, Figure \ref{fig5} shows the distribution of final inclinations $i_{mut}$ of the planetary orbit with respect to that of the binary component. In type-II outcomes (left-hand plots), the secondary remains the original companion $m_B$, therefore $i_{mut}$ measures the perturbation of the fly-by on the original system. Although most orbits remain relatively aligned, especially those corresponding to positive results, we also note a significant amount of highly inclined orbits, some even retrograde with respect to the binary orbit. 

The right-hand plots in Figure \ref{fig5} correspond to type-III outcomes, in which the new binary system is now composed of $m_A+m_C$. Because the new companion is captured and not primordial, in principle there should be little correlation between the orbits of $m_p$ and $m_C$; consequently the values of  $i_{mut}$ should be fairly random. However, the results of the simulations, especially those classified as positive, show a marked correlation. This seems to indicate that only the fly-bys that occur near the orbital plane of the original binary lead to systems suitable for planetary formation. The difference in the plots between the black and red curves gives the impression that most of the encounters with large $I_B$ lead to highly eccentric planets, and positive results occur primarily with  $I_B \sim 0$. 

\begin{table}
\centering
\begin{tabular}{ c | r r r r r }
\hline  \\ [-2ex]
     $m_C$                                        &   $0.2$  &   $0.4$  &   $0.6$   &   $0.8$   & $1.0$ \\ [0.5ex]
\hline \\ [-1ex]
  Type I                                             &       --     &   0.2\% &    0.6 \% &    0.9\% &    2.7\%  \\ [1ex]
  Type II   \\ ($\Delta v < 10$ m/s)    &       --     &   0.1\% &    0.2 \% &    0.2\% &    0.4\% \\ [1ex]
  Type II   \\ ($\Delta v < 100$ m/s)  &    0.9\% &   1.6\% &    2.3 \% &    2.6\% &    5.5\% \\ [1ex]
  Type III  \\ ($\Delta v < 10$ m/s)    &      --      &     --      &      --       &      --      &     --       \\ [1ex]
  Type III  \\ ($\Delta v < 100$ m/s)  &      --      &   0.1\% &    0.1 \% &    0.1\% &    0.1\% \\ [1ex]
\hline
\end{tabular}
\caption{Percentage of positive outcomes per event type and impactor mass for full 3D simulations.}
\label{table2}
\end{table}

\section{Distribution functions for the free parameters}

To gain a more accurate understanding of the probability to form the $\gamma$-Cephei system through the process explained in these paper, we must weight the results showed in the previous sections by assigning a distribution function (DF) to each of the free parameters, $r_{min}$, $v^*$ and $m_C$.

As mentioned earlier, the parameter $r_{min}$ appears more suitable than the impact parameter $b$ for describing the two-body dynamics of the scattering problem. However, to find the corresponding DF, we must first relate $r_{min}$ with $b$ and $v^*$. 

For a hyperbolic keplerian orbit we may write
\begin{eqnarray}
& r_{min}   & = - a (e - 1) \nonumber  \\
& b              & = - a \sqrt{e^2 - 1} \\
& {v^*}^2 & =   \frac{\Lambda}{a}, \nonumber
\label{eq26}
\end{eqnarray}
where $a$ and $e$ are the semimajor axis and eccentricity of the orbit of the incoming body $m_C$ and $\Lambda = a_{AB} \; m_C(m_A+m_B)/(m_A m_B)$. Introducing the last two expressions of (26) into the first, and after some simple algebraic manipulations, we obtain

\eq{
r_{min}(b,v^*) = \frac{\Lambda}{{v^*}^2} \left(\sqrt{\left(\frac{b {v^*}^2}{\Lambda} \right)^2 + 1} - 1 \right).
\label{eq27}
}

Following Hut \& Bahcall (1983) we assumed a distribution of close encounters uniform in $b^2$. To estimate the distribution probability of $v^*$, we first need to determine the DF of the relative encounter velocities between a single star and a binary. Again following Hut \& Bahcall (1983), the thermal distribution function of velocities for stars of equal mass $m$ is
\eq{
f(v^*) = \left( \frac{2}{\pi} \right)^{1/2} \left(\frac{m}{kT} \right)^{3/2} {v^*}^2 \exp{\left(-\frac{m}{2kT}{v^*}^2\right)},
\label{eq28}
}
where $k$ is Boltzmann's constant and $T$ is a temperature-like parameter in analogy with a velocity DF of gas particles. Below we will express these variables in terms of the thermal velocity dispersion $v_{th}$.

Binaries formed from equal-mass stars have a combined mass $2m$, and a thermal DF
\eq{
f'(v^*) = \left( \frac{2}{\pi} \right)^{1/2} \left(\frac{2m}{kT'} \right)^{3/2} {v^*}^2 \exp{\left(-\frac{m}{kT'}{v^*}^2\right)},
\label{eq29}
}
where $T' = T$ for equipartition of translational energy, and $T' = 2T$ if the binaries have the same velocity dispersion as single stars (Hut \& Bahcall 1982). The real value for $T'$ lies somewhere in the middle of these extremes.

Finally, we can calculate the distribution of relative velocities of single stars in the instantaneous rest frame of a binary. It is again a Maxwellian distribution but with an effective mass $m^*$ and temperature $T^*$:
\begin{eqnarray}
m^* &=& \frac{2m^2}{m + 2m} = \frac{2}{3}m \\
T^*  &=& \frac{m^*}{m}T + \frac{m^*}{2m}T' = \frac{2}{3}T + \frac{1}{3}T' \nonumber.
\label{eq30}
\end{eqnarray}
And we have for the velocities
\eq{
f^*(v^*) = \left( \frac{2}{\pi} \right)^{1/2} \left(\frac{m^*}{kT^*} \right)^{3/2} {v^*}^2 \exp{\left(-\frac{m^*}{2kT^*}{v^*}^2\right)}.
\label{eq31}
}

From now on we will characterize the temperature $T^*$ by the thermal velocity dispersion in the relative velocities:
\eq{
s^2 = \left< {v^*}^2 \right> = 3\frac{kT^*}{m^*}.
\label{eq32}
}
Although these expressions were obtained for equal-mass binaries, the same procedure can be extended to other masses. The velocity dispersion is the classical parameter used to characterize the relative velocities of different types of stellar systems. Typical values for $s$ are extremely dependent on the environment, and may be as low as $0.3$ km/s in open clusters, or up to two orders of magnitude higher in the solar neighborhood (see Binney \& Tremaine 2008 for a detailed analysis).

\begin{figure}[t!]  
\centering
\mbox{\includegraphics*[width=8.0cm]{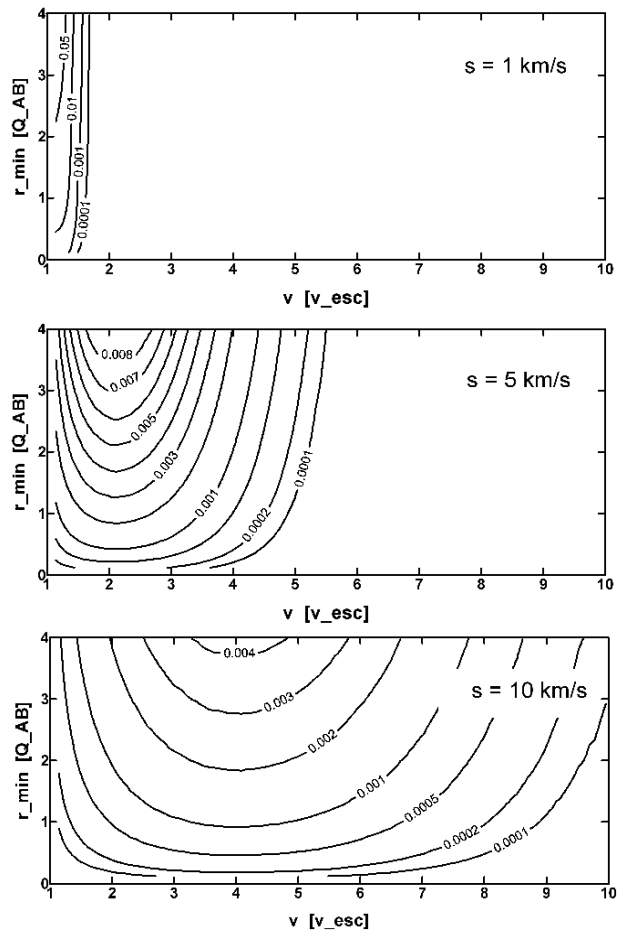}}
\caption{Relative probability function $P(r_{min},v^*) \, dr_{min}\, dv^*$ for stellar encounters with given values of the incoming speed $v$ (in units of $v_c$ and minimum distance $r_{rmin}$ between the impactor and the original binary system (in units of the apocentric distance $Q_{AB}$). In each graph we adopt a different value for the velocity dispersion $s$.}
\label{fig6}
\end{figure}

From equations (26), (\ref{eq27}), (\ref{eq31}) and (\ref{eq32}), we can calculate the probability function $P(r_{min},v^*) \, dr_{min}\, dv^*$ for a given value of $s$. Results are shown in Figure \ref{fig6} for three magnitudes of the velocity dispersion, and adopting values for $dr_{min}$ and $dv^*$ from a $100 \times 100$ grid of initial conditions in the plane.

As expected, higher velocity dispersions increase the probability of encounters at higher values of $v^*$. For $s=1$ km/s, only encounters with $v_0 \sim v_{esc}$ are relevant and most of the solutions discussed in the  lower plot of Figure \ref{fig3} will have negligible weight in the overall outcome probability. On the other hand, for $s=10$ km/s the spread in $v^*$ is much larger and even high incoming speeds are statistically pertinent. 

We can now incorporate these distribution functions into the results shown in Table \ref{table2} to give more realistic probabilities of positive outcomes. Before discussing these results, we must also incorporate a distribution function for the impactor mass. For this we have adopted the so-called universal initial mass function proposed by Kroupa (2001) which, in the mass interval between $\sim 0.1 M_{\odot}$ and $\sim 1 M_{\odot}$, can be expressed by a multiple-part power-law $\xi(m) \propto m^{-\alpha{_i}}$ with
\begin{eqnarray}
\alpha_1 &=& 1.3  \;\;\; {\rm if} \;\;\; 0.08 \le m/M_{\odot} < 0.5 \\
\alpha_2 &=& 2.3  \;\;\; {\rm if} \;\;\; 0.50 \le m/M_{\odot} \nonumber
\label{eq33}
\end{eqnarray}
and where $\xi(m) \, dm$ is the number of single stars expected in the mass interval $[m,m+dm]$. This distribution function is shallower than the classical Salpeter (1955) prescription for small stellar masses but, at least for our application, both give similar results. 

Taking into consideration all distribution functions, we can now estimate the final probability for favorable scattering outcomes, integrated over the impactor mass $m_C$ and initial values for $r_{min}$ and $v^*$. Results are shown in Table \ref{table3} (which adds all positive type I + type II + type III outcomes), where we have considered three different possible values for the velocity dispersion $s$ of background stars. We have also performed separate calculations assuming that the maximum impact speeds for planetesimals leading to accretional collisions is either $10$ m/s or $100$ m/s. We recall that the first is a very strict limit in which all collisions should be constructive (Th\'ebault et al. 2006), while the higher impact velocity allows some disruptions to take place. 

Contrary to expectations, results are not extremely sensitive to the adopted value for $s$. For very stringent accretional restrictions, we find that around $1.5 \%$ of all stellar scattering events lead to favorable outcomes, while this number increases to $\sim 5.5 \%$ if we allow higher impact speeds. This seems to imply that given a ``normal'' planetary system consisting in either (i) a planet around a single star interacting with a binary stellar system, or (ii) a planet in a circumstellar orbit around a wide binary interacting with a single star, the final result could be a system similar to $\gamma$-Cephei in (roughly) $1\% - 5 \%$ of cases.

\begin{table}
\centering
\begin{tabular}{ l | l | r }
\hline
 & & \\
                                  &   $s = 1$ km/s    &   0.9 \% \\ [1ex]
 $\Delta v < 10$ m/s   &   $s = 5$ km/s     &   1.7 \% \\ [1ex]
                                  &   $s = 10$ km/s   &   1.5 \% \\ [1ex]
\hline
 & & \\
                                   &   $s = 1$ km/s    &   3.3 \% \\ [1ex]
$\Delta v < 100$ m/s   &   $s = 5$ km/s     &   5.4 \% \\ [1ex]
                                   &   $s = 10$ km/s   &   5.8 \% \\ [1ex]
\hline
\end{tabular}
\caption{Percentage of positive outcomes for two limiting impact velocities $\Delta v$ for accretion and 
three different velocity dispersions $s$ for background star. Values are integrated over distribution probabilities 
for stellar mass, impact speeds and cross-sections.}
\label{table3}
\end{table}

\section{Conclusions}

The discovery of at least four exoplanets in circumstellar orbits in tight binary systems has raised the question of their formation history. So far, all attempts to explain their origin via normal in-situ formation processes have been unsuccessful, and we wonder whether other more exotic mechanisms may have played a role. Following the suggestion of Marzari \& Barbieri (2007a), in this work we have analyzed whether these systems, in particular $\gamma-Chepei$, could have begun their lives in other dynamical configurations more suitable for planetary formation, attaining their present structure later on as a consequence of a close encounter with a background singleton.

Following a series of numerical simulations of time-reversed stellar encounters, whose results were later weighted adopting classical expressions for probability distributions for stellar mass, relative velocities and impact cross-sections, we have found that between $1$ and $5$ percent of fly-bys between a planetary system and background stars could lead to planets in tight binary systems. Although this number may not seem high, it could in fact explain why planets in tight binaries are not more numerous, which would indeed be the case if in-situ planetary formation were possible. 

However, we must remain cautious. Although we have attempted to give some estimate about the probability outcomes, given a sufficiently large number of initial conditions and values in the parameter space, it is possible to obtain almost any result from a scattering event. This does not mean that the event actually happened or that the explanation lies there. However, as Sir Arthur Conan Doyle said through his character Sherlock Holmes: {\it ``Once you eliminate the impossible, whatever remains, no matter how improbable, must be the truth.''}

\section*{Acknowledgments}
This work has been partially supported by the Argentinian Research Council -CONICET-, and by the Universidad Nacional de C\'ordoba (UNC). The authors wish to express their gratitude to Francesco Marzari for helpful suggestions and a detailed review of this work.

\end{document}